\def\a{\alpha}\def\b{\beta}

\def\l{\lambda}\def\m{\mu}\def\n{\nu}\def\r{\rho}\def\s{\sigma}
\def\y{\eta}

\def\mo{{-1}}\def\ha{{1\over 2}}
\def\qu{{1\over 4}}

\def\({\left(}\def\){\right)}\def\[{\left[}\def\]{\right]}
\def\lra{\leftrightarrow}\def\gdot{{\cdot}}

\def\mn{{\mu\nu}}

\def\coo{coordinates }

\def\poi{Poincar\'e }
\def\des{de Sitter }

\def\SR{special relativity }

\def\cor{commutation relations }

\def\nc{noncommutative }
\def\qft{quantum field theory }

\def\section#1{\bigskip\noindent{\bf#1}\smallskip}

\def\subsection#1{\smallskip\noindent{\it#1}\smallskip}

\def\PL#1{Phys.\ Lett.\ {\bf#1}}\def\CMP#1{Commun.\ Math.\ Phys.\ {\bf#1}}
\def\PRL#1{Phys.\ Rev.\ Lett.\ {\bf#1}}
\def\PR#1{Phys.\ Rev.\ {\bf#1}}\def\CQG#1{Class.\ Quantum Grav.\ {\bf#1}}

 \def\IJMP#1{Int.\ J. Mod.\ Phys.\ {\bf #1}}

\def\JHEP#1{JHEP\ {\bf#1}}\def\JCAP#1{JCAP\ {\bf#1}}\def\EPJ#1{Eur.\ Phys.\ J.\ {\bf#1}}
\def\RMP#1{Rev.\ Mod.\ Phys.\ {\bf#1}}\def\AdP#1{Annalen Phys.\ {\bf#1}}
\def\AHEP#1{Adv.\ High En.\ Phys.\ {\bf#1}}

\def\ref#1{\medskip\everypar={\hangindent 2\parindent}#1}
\def\beginref{\begingroup
\bigskip
\centerline{\bf References}
\nobreak\noindent}
\def\endref{\par\endgroup}

\def\gdot{{\cdot}}

\def\hx{\hat x}\def\hp{\hat p}\def\xp{\,x{\cdot}p\,} \def\px{\,p{\cdot}x\,}
\def\Fd{F^\dagger }\def\Gd{G^\dagger }\def\Hd{H^\dagger }\def\Kd{K^\dagger }

{\nopagenumbers
\line{}
\vskip40pt
\centerline{\bf  Generalizations of Snyder model to curved spaces}
\vskip40pt
\centerline{{\bf S. Meljanac}\footnote{$^\dagger$}{e-mail: meljanac@irb.hr}}
\vskip5pt
\centerline {Rudjer Bo\v skovi\'c Institute, Theoretical Physics Division}
\centerline{Bljeni\v cka c. 54, 10002 Zagreb, Croatia}
\vskip10pt
\centerline{and}
\vskip5pt
\centerline{{\bf S. Mignemi}\footnote{$^\ast$}{e-mail: smignemi@unica.it}}
\vskip5pt
\centerline {Dipartimento di Matematica, Universit\`a di Cagliari}
\centerline{via Ospedale 72, 09124 Cagliari, Italy}
\smallskip
\centerline{and INFN, Sezione di Cagliari}
\centerline{Cittadella Universitaria, 09042 Monserrato, Italy}

\vskip40pt
\centerline{\bf Abstract}
\medskip
{\noindent We consider generalizations of the Snyder algebra to a curved spacetime background with de Sitter symmetry.
As special cases, we obtain the algebras of the Yang model and of triply special relativity.
We discuss the realizations of these algebras in terms of canonical phase space coordinates, up to fourth order in the deformation parameters.
In the case of triply special relativity we also find exact realization, exploiting its algebraic relation with the Snyder model.
}
\vskip60pt
\vfil\eject}

\section{1. Introduction}
Noncommutative geometries have often been advocated as plausible candidates for describing physics at the Planck scale [1].
The first model of noncommutative spacetime was suggested by Snyder [2] in 1947.
Soon after the publication of Snyder's paper, C.N.~Yang proposed a model that combined noncommutativity with spacetime curvature [3].
Yang's model was based on the fifteen-dimensional SO(1,5) algebra. The generators of this algebra were identified with the coordinates
of a phase space with de Sitter symmetry and with the generators of Lorentz transformation. The remaining generator rotates positions into momenta,
but its physical meaning was not specified.

More recently, Kowalski-Glikman and Smolin [4] proposed a model inspired by that of Yang, which realizes the same symmetries in a nonlinear way,
reducing to fourteen the number of generators. They called this model triply special relativity (TSR) because it contains three fundamental constants,
identified with the speed of light, the Planck length and the cosmological constant, generalizing in this way the idea advanced in doubly \SR theories
[5] of deforming the \poi symmetry by the introduction of a new fundamental constant. A particularly interesting property of TSR [6] is that it
realizes the Born duality [7] for the exchange of position and momentum operators.
Another interesting consequence of this model is the prediction of the existence of both a minimal length and a minimal momentum [8].

Later, one of us showed that this model can be realized exactly in terms of coordinates and momenta only [8], and introduced the alternative denomination of
Snyder-de Sitter (SdS) spacetime. In [8,9] it was also shown that TSR algebra can be obtained from the Snyder algebra by a nonunitary transformation.

While we are not aware of other papers dealing with the Yang model, except the recent proposal of a supersymmetric extension of the algebra [10],
a number of articles have investigated aspects of TSR.  Most of them treat its classical limit, either in a nonrelativistic or relativistic setting
[11]. Also the \qft of a self-interacting scalar field in SdS spacetime has been investigated in [12].

It is known that a fruitful approach to \nc geometry is based on Hopf algebras [13], that describe the symmetries of the quantum spacetime.
A powerful tool in this formalism are realizations of Hopf algebras in terms of the Heisenberg algebra, that were introduced in [14-16].
Only recently this approach has been considered in the context of Yang and TSR models in [17]. In this paper, it has been proposed that TSR and a slight
generalization of the Yang model can be treated in a unified way in this formalism.

In the present paper, we discuss general perturbative realizations of the unified model proposed in [17], in terms of the standard Heisenberg algebra.
We also exploit the relation of TSR with the Snyder model to write down some exact realizations. This results should consent to define a
star product and a twist following the approach of [15-17]. This topic is currently being investigated.

\section{2. The model}
We consider a noncommutative algebra of the form [17]
$$[\hx_\m,\hx_\n]=i\b^2M_\mn,\qquad[\hp_\m,\hp_\n]=i\a^2M_\mn,\qquad[\hx_\m,\hp_\n]=ig_\mn,\eqno(1)$$
$$[M_\mn,\hx_\l]=i(\y_{\m\l}\hx_\n-\y_{\n\l}\hx_\m),\qquad[M_\mn,\hp_\l]=i(\y_{\m\l}\hp_\n-\y_{\n\l}\hp_\m),\eqno(2)$$
$$[M_\mn,M_{\r\s}]=i\big(\y_{\m\r}M_{\n\s}-\y_{\m\s}M_{\n\r}-\y_{\n\r}M_{\m\s}+\y_{\n\s}M_{\m\r}\big).\eqno(3)$$
with real parameters $\a$ and $\b$ and $\y_\mn$ the flat metric. We interpret the Hermitian operators $\hx_\m=\hx_\m^\dagger$ and $\hp_\m=\hp_\m^\dagger$ as
\coo of the phase space and $M_\mn=M_\mn^\dagger$ as generators of Lorentz transformations.
The rank-2 tensor $g_\mn$ depends on $\hx_\m$, $\hp_\m$ and $M_\mn$, with $g_\mn^\dagger=g_\mn$.
The algebra (1)-(3) is invariant under Born duality, $\a\lra\b$, $\hx_\m\lra\hp_\m$, $M_\mn\lra M_\mn$, $g_\mn\lra g_{\n\m}$.
In the limit $\b\to0$ it contains  as a subalgebra the de Sitter algebra, in the limit $\a\to0$ the Snyder algebra.

The Jacobi identities imply
$$[M_\mn,g_{\r\s}]=i\big(\y_{\m\r}g_{\n\s}-\y_{\m\s}g_{\n\r}-\y_{\n\r}g_{\m\s}+\y_{\n\s}g_{\m\r}\big),\eqno(4)$$
$$[g_{\l\m},\hx_\n]-[g_{\l\n},\hx_\m]=i(\y_{\m\l}\hp_\n-\y_{\n\l}\hp_\m),\qquad[g_{\l\m},\hp_\n]-[g_{\l\n},\hp_\m]=i(\y_{\m\l}\hx_\n-\y_{\n\l}\hx_\m),\eqno(5)$$
$$[g_\mn,g_{\r\s}]=i\Big([\,[g_\mn,\hp_\s],\hx_\r]-[\,[g_\mn,\hx_\r],\hp_\s]\Big).\eqno(6)$$

Depending on the form of $g_\mn$, one can recover well known models. For example, the Yang model is characterized by the choice [3]
$$g_\mn=h(\hx^2,\hx\gdot\hp+\hp\gdot\hx,\hp^2)\,\y_\mn,\eqno(7)$$
while the TSR model is characterized by [4]
$$g_\mn=\y_\mn+\a^2\hx_\m\hx_\n+\b^2\hp_\m\hp_\n+\a\b(\hx_\m\hp_\n+\hp_\m\hx_\n-M_\mn).\eqno(8)$$

\section{3. Hermitian realizations}
We are interested in finding Hermitian realizations of the above models in phase space, in terms of canonical variables $x_\m$ and $p_\m$,
satisfying $[x_\m,x_\n]=[p_\m,p_\n]=0$, $[x_\m,p_\n]=i\y_\mn$.
In particular, we shall look for representations where the generators $M_\mn$ and $g_\mn$ can be written in terms of $x_\m$ and $p_\m$.
Therefore, we assume
$$\hx_\m=\ha\Big(x_\m F+\Fd x_\m+p_\m G+\Gd p_\m\Big),\qquad\hp_\m=\ha\Big(p_\m H+\Hd p_\m+x_\m K+\Kd x_\m\Big),\eqno(9)$$
$$M_\mn=x_\m p_\n-x_\n p_\m,\eqno(10)$$
$$\eqalign{g_\mn=&\ \y_\mn h_0+x_\m x_\n h_1+h_1^\dagger x_\m x_\n +p_\m p_\n h_2+h_2^\dagger p_\m p_\n+(x_\m p_\n+p_\n x_\m)h_3\cr
&+h_3^\dagger(x_\m p_\n+p_\n x_\m)+(x_\n p_\m+p_\m x_\n)h_4+h_4^\dagger(x_\n p_\m+p_\m x_\n),}\eqno(11)$$
where $F$, $G$, $H$, $K$, $h_i$ are Lorentz-invariant functions of $x^2$, $\xp+\px$ and $p^2$.

In the following, we shall consider these realizations in a perturbative expansion in $\a$ and $\b$.
At second order, we make the ansatz
$$\eqalign{\hx_\m&=x_\m+{a_1\over2}\a\b(x_\m\xp+\px x_\m)+{a_2\over2}\b^2(x_\m p^2+p^2 x_\m)+{a_3\over2}\b^2(p_\m\px+\xp p_\m)+{a_4\over2}\a\b(p_\m x^2+x^2p_\m),\cr
\hp_\m&=p_\m+{b_1\over2}\a\b(p_\m\px+\xp p_\m)+{b_2\over2}\a^2(p_\m x^2+x^2 p_\m)+{b_3\over2}\a^2(x_\m\xp+\px x_\m)+{b_4\over2}\a\b(x_\m p^2+p^2x_\m),}\eqno(12)$$
with constant $a_i$, $b_i$.

We may also go to the next order, with the ansatz
$$\eqalignno{\hx^{(4)}_\m=&\ {c_1\over2}\a^3\b(x_\m x^2\xp+\px x^2x_\m)+{c_2\over2}\a^2\b^2(x_\m x^2p^2+p^2x^2x_\m)+{c_3\over2}\a^2\b^2(x_\m\xp\px+\px\xp x_\m)&\cr
&+{c_4\over2}\a\b^3(x_\m\xp p^2+p^2\px x_\m)+{c_5\over2}\b^4(x_\m p^4+p^4x_\m)+{c_6\over2}\b^4(p_\m\xp p^2+p^2\px p_\m)&\cr
&+{c_7\over2}\a\b^3(p_\m p^2x^2+x^2p^2p_\m)+{c_8\over2}\a\b^3(p_\m\px\xp+\px\xp p_\m)+{c_9\over2}\a^2\b^2(p_\m x^2\xp+\px x^2p_\m)&\cr
&+{c_{10}\over2}\a^3\b(p_\m x^4+x^4p_\m),&(13)}$$
$$\eqalignno{\hp^{(4)}_\m=&\ {d_1\over2}\a\b^3(p_\m p^2\px+\xp p^2p_\m)+{d_2\over2}\a^2\b^2(p_\m p^2x^2+x^2p^2p_\m)+{d_3\over2}\a^2\b^2(p_\m\px\xp+\xp\px p_\m)&\cr
&+{d_4\over2}\a^3\b(p_\m\px x^2+x^2\xp p_\m)+{d_5\over2}\a^4(p_\m x^4+x^4p_\m)+{d_6\over2}\a^4(x_\m\px x^2+x^2\xp x_\m)&\cr
&+{d_7\over2}\a^3\b(x_\m x^2p^2+p^2x^2x_\m)+{d_8\over2}\a^3\b(x_\m\xp\px+\xp\px x_\m)+{d_9\over2}\a^2\b^2(x_\m p^2\px+\xp p^2x_\m)&\cr
&+{d_{10}\over2}\a\b^3(x_\m p^4+p^4x_\m),&(14)}$$
where $c_i$, $d_i$ are constants.
Born dual realizations of (12)-(14) are obtained by $\hx_\m\lra\hp_\m$, $\a\lra\b$ and $x_\m\lra p_\m$.
\bigbreak

\section{4. Yang model}
The original Yang model was characterized by an algebra where $g_\mn$ was considered as an independent generator.
Here we adopt instead the definition (7), where $g_\mn$ is a Hermitian operator, written in terms of a Lorentz-invariant
function $h$ of the phase space variables $\hx_\m$ and $\hp_\m$. Clearly, at zeroth order, $h^{(0)}=1$.
At second order in $\a$ and $\b$ we can set
$$h^{(2)}=g_1\a^2x^2+g_2\a\b(\xp+\px)+g_3\b^2p^2.\eqno(15)$$
with $g_i$ real parameters.
One easily sees that the realization (12) satisfies the Yang algebra if
$$a_2=b_2=-\ha,\quad a_3=b_3=0,\quad a_1+b_1=0\quad a_4+b_4=0,\eqno(16)$$
and
$$h^{(2)}=-\ha\Big(\a^2 x^2+\b^2 p^2\Big).\eqno(17)$$
At this order, the simplest realization of the Yang algebra is given by the choice $a_1=b_1=a_4=b_4=0$.

To fourth order, we can assume

$$\eqalign{h^{(4)}&=e_1\a^4x^4+{e_2\over2}\a^3\b(x^2\xp+\px x^2)+{e_3\over2}\a^2\b^2(x^2p^2+p^2x^2)\cr
&+e_4\a^2\b^2\xp\px+{e_5\over2}\a\b^3(\xp p^2+p^2\px)+e_6\b^4p^4.}\eqno(18)$$

Inserting (13)-(14) in the Yang algebra, one gets the independent conditions
$$c_6=d_6=0,\quad c_6-4c_5=\ha,\quad d_6-4d_5=\ha,\quad c_8+d_1=0,\quad d_8+c_1=0,$$
$$c_1+d_4=a_1b_2+2a_4b_2,\quad d_1+c_4=a_2b_1+2a_2b_4,\quad 2c_2+d_9=-a_1b_4,\quad 2d_2+c_9=-a_4b_1,$$
$$c_7+2d_{10}=-a_2b_4,\quad d_7+2c_{10}=-a_4b_2,\quad c_3+d_3=-a_1b_1+2a_2b_2,\quad c_9+d_9=-a_4b_1-a_1b_4-2a_4b_4.$$
with
$$e_1=d_5,\quad e_2=c_1+d_4+a_1b_2,\quad e_3=c_2+d_2+a_2b_2-a_4b_4,$$
$$e_4=c_3+d_3+a_1b_1,\quad e_5=c_4+d_1+a_2b_1,\quad e_6=c_5.$$
Taking into account (16), it follows that at this order there are six further independent parameters
$c_1$, $c_2$, $c_3$, $c_4$, $c_7$ and $c_{10}$, with
$$c_6=d_6=0,\quad c_5=d_5=-{1\over8},\quad c_8=-{a_1\over2}-a_4+c_4,\quad c_9=a_1a_4+2a_4^2+2c_2,$$
$$d_1={a_1\over2}+a_4-c_4,\quad d_2=-a_4^2-c_2,\quad d_3=\ha+a_1^2-c_3,\quad d_4=-{a_1\over2}-a_4-c_1,$$
$$d_7={a_4\over2}-2c_{10},\quad d_8=-c_1,\quad d_9=a_1a_4-2c_2,\quad d_{10}=-{a_4\over4}-{c_7\over2}.$$
and
$$e_1=e_6=-{1\over8},\quad e_3=\qu,\quad e_4=\ha,\quad e_5=-e_2=a_1+a_4.$$

The simplest Hermitian realization of the Yang model up to fourth order in $\a$, $\b$ is therefore
$$\eqalignno{\hx_\m&=x_\m-{\b^2\over4}\Big(x_\m p^2+p^2 x_\m\Big)-{\b^4\over16}\Big(x_\m p^4+p^4x_\m\Big)+{\a^2\b^2\over8}\Big(x_\m\xp\px+\px\xp x_\m\Big),&\cr
\hp_\m&=p_\m-{\a^2\over4}\Big(p_\m x^2+x^2 p_\m\Big)-{\a^4\over16}\Big(p_\m x^4+x^4p_\m\Big)+{\a^2\b^2\over8}\Big(p_\m\px\xp+\xp\px p_\m\Big).&(19)}$$
with
$$h=1-\ha\Big(\a^2x^2+\b^2p^2\Big)-{1\over8}\Big(\a^2x^2-\b^2p^2\Big)^2+{\a^2\b^2\over2}\,\xp\px.\eqno(20)$$

Note that $h$, $\hx_\m$ and $\hp_\m$ satisfy
$[h,\hx_\m] =i\b^2\hp_\m$ and $[h,\hp_\m] =-i\a^2\hx_\m$.

\section{5. Triply special relativity}
The TSR algebra is defined by (1)-(3) with $g_\mn$ given by (8).
An important relation following from (8) is
$$g_\mn-g_{\n\m}=-2\a\b M_\mn.\eqno(21)$$
From this, after some manipulations, one can obtain an equivalent form of the algebra,
written explicitly in terms of $\hx_\m$ and $\hp_\m$ only, that was first proposed in [8,9].
In those papers, the Lorentz generators were defined as
$$M_\mn=\ha\Big(\hx_\m\hp_\n+\hp_\n\hx_\m-\hx_\n\hp_\m-\hp_\m\hx_\n\Big),\eqno(22)$$
or equivalently, using (21),
$$M_\mn={1\over1-i\a\b}(\hx_\m\hp_\n-\hx_\n\hp_\m)={1\over1+i\a\b}(\hp_\n\hx_\m-\hp_\m\hx_\n).\eqno(23)$$
It follows that $g_\mn$ can be written as
$$g_\mn=\y_\mn+\a^2\hx_\m\hx_\n+\b^2\hp_\n\hp_\m+\a\b(\hx_\n\hp_\m+\hp_\m\hx_\n).\eqno(24)$$

We call this SdS realization of the TSR algebra. Using (22) and (24) we can obtain second-order realizations of the SdS algebra in terms of the
canonical Heisenberg algebra by inserting (10) and (12) in the defining relations.
It is easy to see that the SdS algebra is satisfied if
$$a_1+b_1=0,\quad a_2=b_2=0,\quad a_3=b_3=1,\quad a_4+b_4=1.\eqno(25)$$
The simplest choice of coefficients satisfying these relations is given by $a_1=b_1=0$, $a_4=b_4=\ha$.

At higher order, using the ansatz (13)-(14), one obtains the following independent relations among the parameters:
$$c_5=c_6=d_5=d_6=0,\quad c_2+d_2=a_4b_4,\quad c_3+d_3=a_3b_3-a_1b_1$$
$$c_1+d_4=a_4b_3,\quad d_1+c_4= a_3b_4,\quad c_1+d_8=a_1+b_3,\quad d_1+c_8=b_1+a_3,$$
$$ 2c_7+4d_{10}=b_4-a_3b_4,\quad 2d_7+4c_{10}=a_4-a_4b_3,\quad 2c_2+d_9=2b_4-a_1b_4,\quad 2d_2+c_9=2a_4-a_4b_1.$$

It follows that, taking into account (25), one has, in analogy with the Yang model, six new independent parameters,
say $c_1$, $c_2$, $c_3$, $c_4$, $c_7$, $c_{10}$ and the relations
$$c_5=c_6=0,\quad c_8=-a_1+a_4+c_4,\quad c_9=a_1a_4+2a_4^2+2c_2,\quad d_1=1-a_4-c_4,$$
$$d_2=a_4-a_4^2-c_2,\quad d_3=1-a_1b_1-c_3,\quad d_4=a_4-c_1,\quad d_5=d_6=0,$$
$$d_7=-2c_{10},\quad d_8=1+a_1-c_1,\quad d_9=2-a_1-2a_4+a_1a_4-2c_2,\quad d_{10}=-{c_7\over 2}.$$

A simple realization of the SdS algebra up to fourth order with symmetric $\hx$ and $\hp$ is therefore given by
$$\eqalignno{\hx_\m=&\ x_\m+\Big({\b^2\over2}p_\m\px+{\a\b\over4}p_\m x^2+{\a^2\b^2\over16}x_\m x^2p^2
+{\a^2\b^2\over4}x_\m\xp\px+{\a\b^3\over4}x_\m\xp p^2&\cr&+{\a\b^3\over2}p_\m\px\xp
+{3\a^2\b^2\over8}p_\m x^2\xp+{\rm h.c.}\Big)&(26)\cr
\hp_\m=&\ p_\m+\Big({\a^2\over2}x_\m\xp+{\a\b\over4}x_\m p^2+{\a^2\b^2\over16}p_\m p^2x^2
+{\a^2\b^2\over4}p_\m\px\xp+{\a^3\b\over4}p_\m\px x^2&\cr&+{\a^3\b\over2}x_\m\xp\px
+{3\a^2\b^2\over8}x_\m p^2\px+{\rm h.c.}\Big)&(27)}$$

\section{6. Further developments}
The relation of the TSR algebra with the Snyder algebra was first noticed in [8,9]. Here we exploit it using a different
derivation.
We proceed as follows: the expression (8) of $g_\mn$ can be written as
$$g_\mn=\y_\mn+(\a\hx_\m+\b\hp_\m)(\a\hx_\n+\b\hp_\n) -\a\b M_\mn=\y_\mn+\b^2P_\m P_\n-\a\b M_\mn,\eqno(28)$$
where we have defined $P_\m=\hp_\m+{\a\over\b}\hx_\m$.
Since $\hx^\dagger_\m=\hx_\m$, $\hp^\dagger_\m=\hp_\m$, $M^\dagger_\mn=M_\mn$, it follows that $P^\dagger_\m=P_\m$, $g^\dagger_\mn=g_\mn$,
and consequently from (28) $[P_\m,P_\n]=0$. Hence,
$$[\hx_\m,P_\n]=i(\y_\mn+\b^2P_\m P_\n).\eqno(29)$$
Together with $[\hx_\m,\hx_\n]=i\b^2M_\mn$ this recalls the \cor of the Snyder model [2]. One can hence derive a realization
of the SdS algebra from the realizations of the Snyder model discussed in [15,18], in terms of canonical variables $X_\m$, $P_\n$
satisfying $[X_\m,X_\n]=[P_\m,P_\n]=0$, $[X_\m,P_\n]=i\y_\mn$, namely,
$$\eqalignno{\hx_\m&=X_\m+{\b^2\over2}\Big(X\gdot P\,P_\m+P_\m\,P\gdot X\Big),&\cr
\hp_\m&=P_\m-{\a\over\b}\hx_\m=P_\m-{\a\b\over2}\left(X\gdot PP_\m+P_\m P\gdot X\right)-{\a\over\b}X_\m.&(30)}$$
This realization is not included in the class investigated in the previous section, because $\hp_\m$ contains terms proportional to
$X_\m$ that were neglected in the ansatz (12), but is exact, although it is not symmetric in $X$ and $P$ and is not well defined in
the  limit $\b\to0$.

A realization that is regular for vanishing $\b$, but singular for $\a\to0$ can be obtained by duality, starting from a
representation of \des algebra in Beltrami \coo [19]. One has
$$\eqalignno{\hp_\m&=P_\m+{\a^2\over2}\Big(P\gdot X\,X_\m+X_\m\,X\gdot P\Big),&\cr
\hx_\m&=X_\m-{\b\over\a}\hp_\m=X_\m-{\b\a\over2}\left(P\gdot XX_\m+X_\m X\gdot P\right)-{\b\over\a}P_\m,&(31)}$$
where $X_\m$ and $P_\m$ still satisfy canonical commutation relations, but are not the same as in (30).
\bigskip
Further realizations can be obtained by similarity transformations starting from the ones found above.
In fact, let us consider the \cor (1)-(3)
and act on them with a unitary operator $S$ from the left and $S^\mo$ from the right, defining
$$\hx'_\m=S\hx_\m S^\mo,\qquad\hp'_\m=S\hp_\m S^\mo,\qquad M'_\mn=SM_\mn S^\mo=M_\mn,\qquad g'_\mn=Sg_\mn S^\mo.\eqno(32)$$
Then $\hx'_\m$ and $\hp'_\m$ satisfy the same \cor as $\hx_\m$ and $\hp_\m$.

We may write $S=e^{iG}$, with $G=G(x^2,\xp\!+\!\px,p^2)$, $G^\dagger=G$ and $[M_\mn,G]=0$.
In this way we generate infinitely many realizations of $\hx$ and $\hp$ in terms of $x$ and $p$, satisfying the same algebra.

\section{7. Conclusions}
We have discussed a general quantum algebra that depends on two parameters $\a$ and $\b$, usually identified with a minimum
length and the cosmological constant, and includes as special cases Yang [3] and TSR [4] algebras.
This algebra is relevant for quantum gravity research, because it combines the effects of non- commutativity with those of the
curvature of spacetime, a subject that has attracted a large interest recently [21].

We have found realizations of these quantum algebras on canonical phase space.
The form of the algebra (1)-(3) is much more general than the special cases we have considered, and we are now constructing
more general models of this class. In particular, even considering algebras not more than quadratic in the generators, several
possibilities are available.

A more difficult problem is to construct a quasi-Hopf algebra associated to these models. It seems that star product and twist have not been
constructed so far for Yang and TSR models. Star products should be nonassociative as for the Snyder model [15,17,18].
Star products, related to noncommutative coordinates $\hx_\m$ whose realizations depend on the parameters $\a$ and $\b$ and on
$x_\m$, $p_\m$ are under construction, exploiting the method proposed in [15,18,20].
This construction implies a generalization of the Hopf algebroid approach [22].

An interesting field of application of our results is QFT. A field theory based on the SdS algebra has been
discussed in [12], where it was also remarked its similitude with the Grosse-Wulkenhaar model [23]. This model is of primary relevance
because it gives rise to a renormalizable and exactly solvable theory, which, in analogy with SdS field theory, can be thought as a field theory
in noncommutative curved space [24].

\section{Ackowledgements}
We wish to thank J. Lukierski and M. Woronowicz for interesting discussions. S. Mignemi acknowledges support from GNFM and COST action CA18108.

\beginref
\ref [1] S. Doplicher, K. Fredenhagen and J. E. Roberts, \PL{B331}, 39 (1994).
\ref [2] H.S. Snyder, \PR{71}, 38 (1947).
\ref [3] C.N. Yang, \PR{72}, 874 (1947).
\ref [4] J. Kowalski-Glikman and L. Smolin, \PR{D70}, 065020 (2004).
\ref [5] G. Amelino-Camelia, \PL{B510}, 255 (2001). J. Magueijo and L. Smolin, \PRL{88}, 190403 (2002).
\ref [6] H.G. Guo, C.G. Huang and H.T. Wu, \PL{B663}, 270 (2008).
\ref [7] M. Born, \RMP{21}, 463 (1949).
\ref [8] S. Mignemi, \CQG{29}, 215019 (2012).
\ref [9] S. Mignemi and R. \v Strajn, \AHEP{2016}, 1328284 (2016).
\ref [10] J. Lukierski and M. Woronowicz, \PL{B824}, 136783 (2021).
\ref [11] C. Chryssomakolos and E. Okon, \IJMP{D13} 1817.
A. Das and O.C.W. Kong, \PR{D 73},124029 (2006).
S. Mignemi, \CQG{26}, 245020 (2009).
M.C. Carrisi and S. Mignemi, \PR{D82}, 105031 (2010).
R. Banerjee, K. Kumar and D. Roychowdhury, \JHEP{1103}, 060 (2011).
B. Iveti\'c, S. Meljanac and S. Mignemi, \CQG{31}, 105010 (2014).
\ref [12]  A. Franchino-Vi\~nas and S. Mignemi, \EPJ{C80}, 382 (2020).
\ref [13] S. Majid,  {\it Foundations of quantum group theory}, Cambridge University Press 1995.
\ref [14] M.V. Battisti and S. Meljanac, \PR{D79}, 067505 (2009).
\ref [15] M.V. Battisti and S. Meljanac, \PR{D82}, 024028 (2010).
\ref [16] S. Meljanac, D. Meljanac, A. Samsarov and M. Stoji\'c, \PR{D83}, 065009 (2011).
T. Juri\'c, S. Meljanac, D. Pikuti\'c and R. \v Strajn, \JHEP{07}, 055 (2015).
\ref [17] S. Meljanac and R. \v Strajn, SIGMA {\bf18}, 022 (2022).
\ref [18] S. Meljanac, D. Meljanac, S. Mignemi and R. \v Strajn, \PL{B768}, 321 (2017).
S. Meljanac, D. Meljanac, F. Mercati and D. Pikuti\'c, \PL{B766}, 1815 (2017).
\ref [19] S. Mignemi, \AdP{522}, 924 (2010).
\ref [20] S. Meljanac, D. Meljanac, S. Mignemi, D. Pikuti\'c and R. \v Strajn, \EPJ{C78}, 194 (2018).
S. Meljanac, D. Meljanac, S. Mignemi and R. \v Strajn, \IJMP{A32}, 1750172 (2017).
\ref [21] G. Rosati, G. Amelino-Camelia, A. Marciano and M. Matassa, \PR{D92}, 124042 (2015).
P. Aschieri, A. Borowiec and A. Pachol, \JCAP{04}, 025 (2021).
A. Ballesteros, G. Gubitosi and F. Mercati, Symmetry {\bf 13}, 2099 (2021).
\ref [22] T. Juric, S. Meljanac and R. Strajn, \PL{A377}, 2472 (2013).
T. Juric, D. Kovacevic and S. Meljanac, SIGMA {\bf 10}, 106 (2014).
J. Lukierski, D. Meljanac, S. Meljanac, D. Pikutic and M. Woronowicz, \PL{B777}, 1 (2018).
\ref [23] H. Grosse and R. Wulkenhaar, \EPJ{C35}, 277 (2004).
H. Grosse and R. Wulkenhaar, \CMP{256}, 305 (2005).
\ref [24] M. Buri\'c and M. Wohlgennant, \JHEP{1003}, 053 (2010).

\endref

\end